# Comments on the Radial Distribution Functions and Structure Factors of Aggregates: Fractal and Non−Fractal Approaches

M. Cattani[(*)], M.C.Salvadori and F.S.Teixeira
Instituto de Fisica, Universidade de S. Paulo, C.P. 66318, CEP 05315−970
S. Paulo, S.P. Brazil . [(*)]E−mail: mcattani@if.usp.br

Abstract

The definitions and applications of Radial Distribution Function (RDF) and Structure Factor (SF) to study properties of aggregate are found in many papers and books. The approach adopted to calculate the RDF and the SF to determine the interaction potential between particles of an aggregate is very different from that adopted to obtain their fractal properties. The difference between the two approaches will be shown in details here. This article was written to graduate and postgraduate students of Physics.

Resumo

As definições e aplicações da Função de Distribuição Radial (FDR) e do Fator de Estrutura (FE) para estudar as propriedades de agregados são encontradas em muitos artigos e livros. A aproximação adotada para calcular a FDR e a FE para determinar o potencial de interação entre as partículas de um agregado é muito diferente daquela que é adotada para obter suas propriedades fractais. A diferença entre os dois métodos será mostrada em detalhes aqui. Esse artigo foi escrito para alunos de graduação e pós−graduação de Física.

## 1. Função de Correlação de Duas Partículas.

Para tornar a análise mais simples suporemos que o material seja composto por N "partículas" esféricas com raio $r_o$ ocupando um volume V. Essas "partículas" ( ou "monômeros") podem ser átomos, moléculas ou também aglomerados nanométricos de átomos ou moléculas. Além disso, assumiremos que a distribuição das partículas nos aglomerados possua simetria esférica. Em Mecânica Estatística, a partir da *"função de correlação de 2 partículas"* ou *"pair correlation function"* $n^{(2)}(r)$ obtém−se a *Função de Distribuição Radial* (FDR) g(r) dada por:[1,2,3]

$$g(r) = (V/N)^2 \, n^{(2)}(r) \qquad (1.1).$$

A g(r), que é normalizada a 1 para $r \rightarrow \infty$, mostra a distribuição efetiva de partículas localizadas a uma distância r em torno de uma outra partícula situada na origem. O número de partículas dN(r) entre r e r+dr é dado por dN(r) = 4π (N/V) r²dr. Pondo Φ = (N/V), o número de partículas N(r) que está dentro de uma esfera de raio r seria dado por

$$N(r) = \Phi \int_0^r g(r)\, 4\pi r^2\, dr \qquad (1.2).$$

## 2. Determinação do Potencial de Interação entre as Partículas.

Com o intuito de determinar o potencial de interação entre as partículas em um material a função g(r) é muitas veses[1,2] escrita na forma

$$g(r) = \exp\{-\varphi(r)/kT\} \qquad (2.1),$$

onde k é constante de Boltzmann, T a temperatura absoluta do sistema e φ(r) leva em conta a energia potencial efetiva de interação entre duas partículas. Na Figura 1 vemos uma função g(r) típica para um líquido monoatômico assumindo que φ(r) é um potencial do tipo, por exemplo, Lennard–Jones.[1,2] O comportamento de g(r) mostrando como estão concentradas as partículas em função de r pode ser facilmente entendido levando em conta[2] que φ(r) para r → 0 descreve uma força repulsiva ("caroço duro"), para r ~ σ = 2$r_o$ = diâmetro da partícula ela passa por um mínimo negativo (devido a uma força atrativa entre as partículas) e para r → ∞ tende a 0. Assim, g(r) = 0 para r ≤ σ = 2$r_o$, passa por um máximo em r ~ σ, na camada dos primeiros vizinhos, e para r > σ passa por oscilações representando o efeito dos vizinhos mais distantes. As amplitudes das oscilações decrescem à medida que r cresce e para r → ∞, onde a distribuição das partículas (densidade) fica uniforme, g(r) →1.

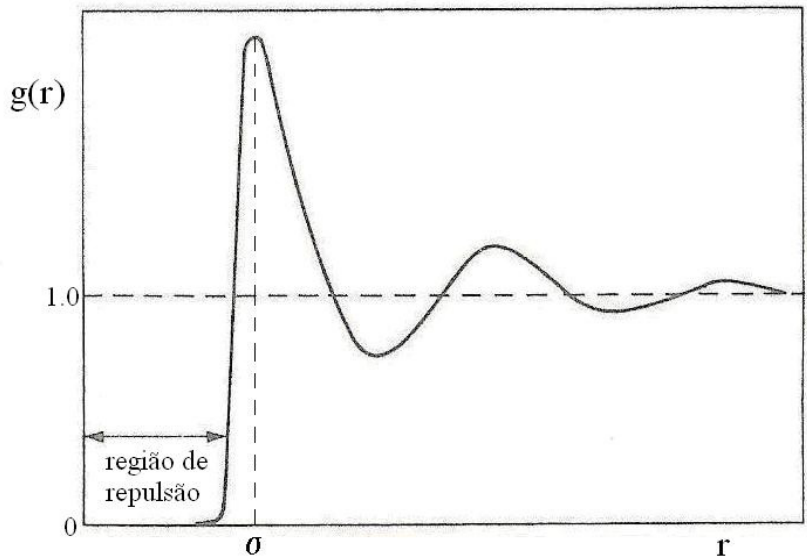

**Figura 1**. Forma típica da função g(r) x r para um líquido denso.[2] O maior pico de g(r) ocorre para a distância r ~ σ = diâmetro da partícula.

Para se determinar as propriedades microscópicas de um material mede−se, por exemplo, a intensidade I(q) de espalhamento[3-5] de nêutrons, elétrons ou raios−X (fotons) pelo material. Ela é dada por $I(q) = I_o(q) S(q)$ onde $I_o(q)$ descreve o espalhamento incoerente devido aos N monômeros e S(q) é o *fator de estrutura* (FE) definido por[1,2,3-5]

$$S(q) = 1 + 4\pi\Phi \int_0^\infty [g(r) - 1] r^2 (\sin(qr)/qr) \, dr \}  \qquad (2.2),$$

onde q é a variação do momento linear do fóton, nêutron ou eletron espalhado. No caso de uma colisão elástica $q = 2k \sin(\theta/2)$ onde $k = 2\pi/\lambda$ é o vetor de onda da partícula incidente e $\theta$ é ângulo de espalhamento. De acordo com a relação quântica de incerteza de posição−momento $\Delta p \, \Delta r \geq \hbar$ vemos que se no espalhamento ocorre uma mudança de momento $\Delta p = \hbar q$ a região que gera tal efeito deve ter um raio $R \sim 1/q$. Como, em geral, a função $\varphi(r)$ é diferente de zero[2] no intervalo $0 < r \leq 10$ Å = 1 nm os detalhes da função $\varphi(r)$ só podem ser obtidos se observarmos espalhamentos com $1/q < 10$ Å = 1 nm, ou seja, as trocas de momento q devem ser $q > 1$ nm$^{-1}$. Se os nêutrons ou fótons incidentes tiverem $\lambda \sim 1$ Å vemos que $q = (4\pi/\lambda) \sin(\theta/2) \sim 4\pi \sin(\theta/2)$. Para que $q > 1$ nm$^{-1}$ o ângulo de desvio $\theta$ deverá ser $\theta > 10°$, ou seja, os espalhamentos efetivos devem ocorrer somente a grandes ângulos.

Na Figura 2 mostramos uma função típica de S(q) em função de q para um líquido monoatômico no caso de *espalhamento a grandes ângulos*. O valor da função S(q) no ponto zero é dado, conforme mostramos no Apêndice, por $S(0) = \Phi \chi_T kT$, onde $\chi_T = (\partial \Phi / \partial P)_{V,T} / \Phi$ é a *compressibilidade isotérmica* do material.[2]

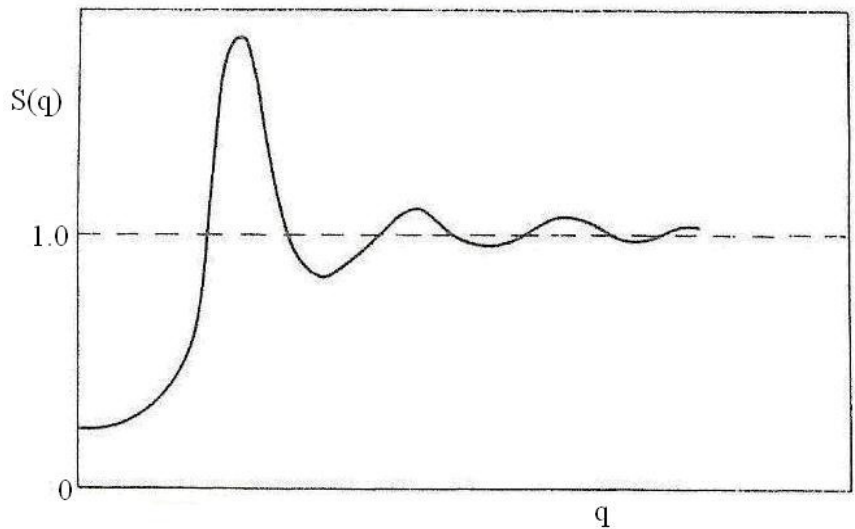

**Figura 2**. Forma típica da função S(q) x q no caso de *espalhamento a grandes ângulos* para um líquido denso[2] em função de q que é medido usualmente em nm$^{-1}$ ou Å$^{-1}$.

Na figura 3 vemos S(q) em função dos grandes ângulos de espalhamento θ de nêutrons no caso do Ar líquido[6] a 84 K.

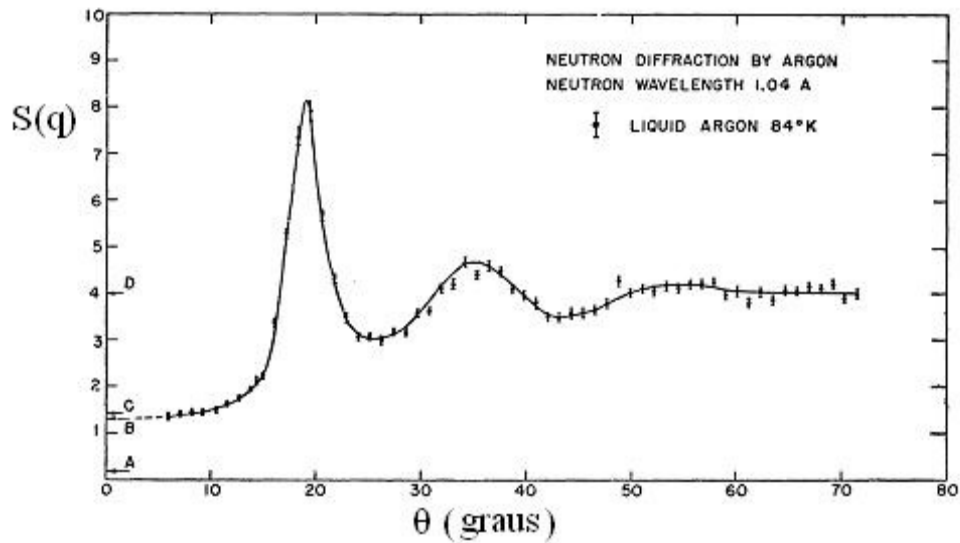

**Figura 3**. S(q) x θ medido em graus, obtida por *espalhamento a grandes ângulos* de nêutrons em Ar líquido[6] a 84 K.

A partir de S(q) determina−se g(r) usando a (2.2). Na Figura 4 vemos os valores experimentais[2,7] de S(q) em função de q para o Rb a 40°C, obtidos por *espalhamento a grandes ângulos* de nêutrons, comparado com os valores de S(q) obtidos com a (2.2) assumindo g(r) gerada por um potencial φ(r) de "esferas duras"[1,2,7]

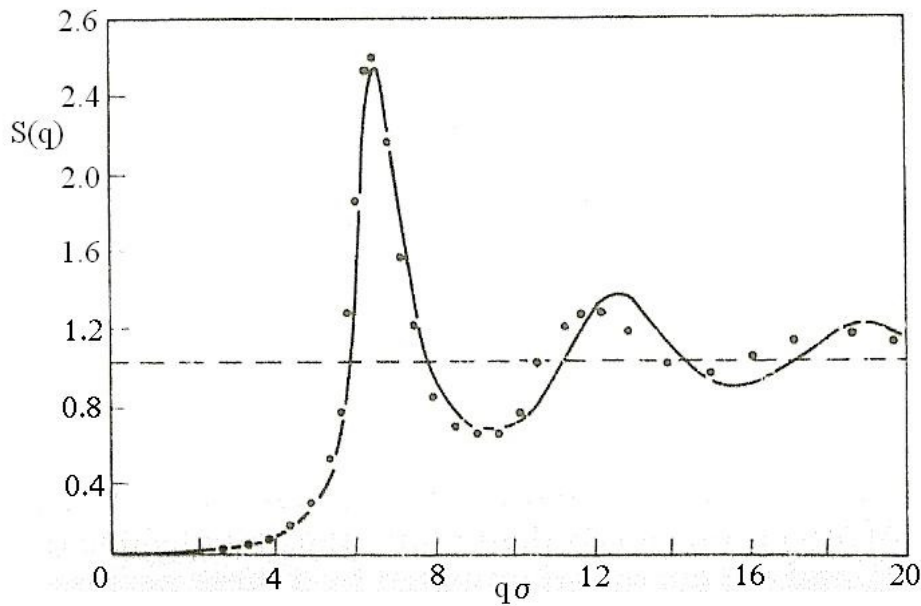

**Figura 4.** Função S(q) x qσ obtida experimentalmente[2] por *espalhamento a grandes ângulos* de nêutrons (representada por bolinhas pretas) para o Rb a 40° C comparada com cálculos (linha contínua) efetuados supondo interação de "esferas duras"[1,2.7] entre as partículas.

Na Figura 5 mostramos g(r) calculada a partir de valores experimentais[2,6] de S(q) obtida por difração a grandes ângulos de nêutrons para o Ar líquido a 84 K supondo φ(r) um potencial Lennard–Jones[1,2,6].

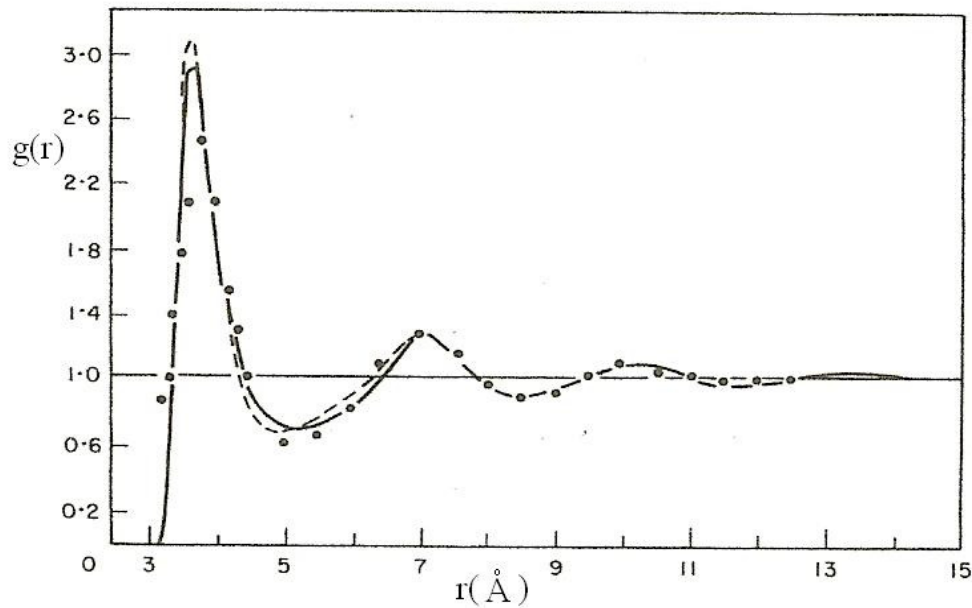

**Figura 5.** Função g(r) x r obtida a partir de S(q) para Ar a 84 K medida[6] por *difração a grandes ângulos* de nêutrons (bolinhas pretas) comparada com g(r) calculada (linhas contínua e tracejada) considerando φ(r) um potencial Lennard–Jones[1,2,6,8].

## 3. Determinação de Propriedades Fractais de Agregados.

Veremos nesta Seção como determinar as propriedades fractais de agregados analisando o espalhamento de nêutrons, elétrons e raios–X (fótons). Para isso vamos antes mostrar como calcular a fractalidade de volume e de massa de sistemas formados por monômeros esféricos com raio $r_o$. Para um estudo mais amplo e geral sobre fractalidade sugerimos a leitura, por exemplo, dos livros de Mandelbrot[9] e Feder.[10]

**(3.a)*Raio da esfera e Número de monômeros.***

Consideremos uma esfera (ou circunferência) ocupada por monômeros ("bolinhas") de raios $r_o$ com um raio dado por $R_n = a^n r_o$ onde n = 1,2,... . Vamos assumir que o número de bolinhas seja dado por $N_n = b^n$. Ou seja,

$$R_n = a^n r_o \quad \text{e} \quad N_n = b^n \quad (n = 1,2,3,...) \quad (3.1).$$

Os nossos cálculos são válidos somente quando $R_n \gg r_o$, ou seja, quando precisarmos de um número $N_n$ muito grande de bolinhas para enchermos completamente o volume da esfera com raio $R_n$.

A (3.1) também pode ser escrita na forma

$$R_n/r_o = a^n \quad e \quad N_n = b^n \qquad (3.2),$$

De onde tiramos (omitindo o índice n, por simplicidade)

$$n = \ln(R/r_o)/\ln(a) = \ln(N)/\ln(b) \qquad (3.3),$$

ou seja,

$$\ln(b)/\ln(a) = D = \ln(N)/\ln(R/r_o) \qquad (3.4).$$

A (3.4) mostra que

$$N = (R/r_o)^D, \text{ onde } D = \ln(b)/\ln(a) \qquad (3.5).$$

**(3.b) *Volume Ocupado ($V_o$) e Volume Disponível ($V_n$).***

O volume ocupado $V_o(n)$ por n bolinhas, cada uma com volume $v_o = (4/3)\pi r_o^3$ e o volume disponível $V(n)$ são dados, respectivamente, por

$$V_o(n) = N_n v_o = b^n v_o$$

e
$$V(n) = (4/3)\pi R_n^3 = (4/3)\pi (a^n r_o)^3 = a^{3n} v_o \qquad (3.6)$$

Devemos esperar que $V(n) \geq V_o(n)$ de onde tiramos, usando (3.6)

$$a^{3n} \geq b^n, \text{ ou seja, } 3n \ln(a) \geq n \ln(b).$$

que dá, usando a (3.5), que $D \leq 3$.

**(3.c) *Volume Ocupado ($V_o$) = Volume Disponível ($V_n$).***

Vejamos qual a condição que deve ser obedecida para que tenhamos $V(n) = V_o(n)$. Ora, para que isto seja válido devemos ter, usando a (3.6):

$$V(n)/V_o(n) = 1 = (b/a^3)^n \qquad (3.7),$$

ou seja, devemos ter $b = a^3$, de onde obtemos usando a (3.5)

$$D = \ln(b)/\ln(a) = 3$$

e, consequentemente, omitindo por simplicidade o índice n, que

$$N = (R/r_o)^3 \qquad (3.8)$$

que é o caso **Euclideano**, ou seja, **D = 3**.

**(3.d)** *Fractalidade de Massa e de Densidade de Massa.*

Sendo $m_o$ a massa de cada monômero e usando a (3.5) a massa M = M(R) do sistema é dada por,

$$M(R) = N\,m_o = m_o(R/r_o)^D \qquad (3.9)$$

e a sua densidade $\rho = M/[(4/3)\pi R^3]$ é dada por

$$\rho(R) = (3m_o/4\pi r_o^D)(R/r_o)^{D-3} \sim (R/r_o)^{D-3} \qquad (3.10).$$

**(3.e)** *Caso Genérico como função da posição r.*

Consideremos uma partícula no ponto r = 0. O número de partículas N(r), a massa M(r) e a densidade $\rho(r)$ dentro de uma distância r serão dadas, respectivamente, por

$$N(r) = (r/r_o)^D, \quad M(r) = m_o(r/r_o)^D \quad \text{e} \quad \rho(r) = (3m_o/4\pi r_o^D)(r/r_o)^{D-3} \quad (3.11).$$

A densidade $\rho(r)$ pode ser escrita como

$$\rho(r) = (3m_o/4\pi r_o^D)(r/r_o)^{D-3} = C\,(r/r_o)^{D-3} \qquad (3.12),$$

onde o parâmetro $C = (3m_o/4\pi r_o^D)$ depende da massa, do raio das partículas e da fractalidade D do meio onde as partículas estão imersas

**(3.f)** *Função de Correlação para um Aglomerado Fractal.*

Vejamos como calcular g(r) e S(r) para um agregado fractal com *dimensão fractal de massa* D. Devemos lembrar que as Eqs.(3.11) para N(r), M(r) e $\rho(r)$ são válidas somente quando r >> $r_o$ e, consequentemente, V >> $v_o$. Desse modo, diferenciando a (1.2) e (3.11) obtemos, respectivamente,

$$dN(r) = \Phi g(r) 4\pi r^2\,dr \quad \text{e} \quad dN(r) = (D/r_o^D)\,r^{D-1}\,dr \qquad (3.13).$$

Igualando os dois resultados teremos,

$$\Phi g(r) = (D/4\pi r_o^D)\,r^{D-3} \qquad (3.14).$$

Como[9,10] D < 3, conforme (3.14), g(r) tende a zero para r → ∞ que é um resultado não físico pois para grandes valores de r o agregado mostra uma densidade macroscópica, uniforme, com flutuações desprezíveis. Como sabemos, conforme a teoria de líquidos,[1,2] g(r) → 1 when r → ∞. Para descrever corretamente o comportamento de um agregado para grandes distâncias introduziu−se[11,12] uma distância de corte $\xi$, denominada de "*distância de correlação*" obtendo:

$$\Phi[g(r) - 1] = (D/4\pi r_o^D)\, r^{D-3} \exp(-r/\xi) \qquad (3.15).$$

Como as relações fractais (3.11)–(3.15) são válidas para r >> $r_o$ a função fractal (3.15) $\Phi[g(r) - 1]$ é completamente diferente da que é obtida usando g(r) = exp{−φ(r)/kT} que, em princípio, é válida para qualquer valor de r. Consequentemente, as correspondentes funções de estrutura S(q) que dependem de $\Phi[g(r)-1]$ são também completamente diferentes. O significado físico de $\xi$ é somente qualitativo e só pode ser precisado analisando cada caso em particular. Em geral ele representa uma distância característica maior do que a qual a distribuição de massa em um cluster não é mais descrita por uma lei fractal. Ele desempenha um papel semelhante à distância de correlação associada com a rugosidade de superfícies de materiais desordenados.[13,14]

Como as relações fractais (3.11)–(3.15) foram deduzidas assumindo que r >> $r_o$ conclui−se que elas não poderão dar informações muito precisas de propriedades fractais do material para distâncias r da ordem de grandeza de $r_o$. Assim, se buscamos determinar com boa precisão as propriedades fractais de um material devemos analisar regiões com tamanhos R > $r_o$. Usando difração de nêutrons ou de raios−X levando em conta que R ~ 1/q devemos vemos que os espalhamentos devem obedecer à condição R ~ 1/q > $r_o$, isto é, que 1 > q $r_o$. Ou seja, a relação (4π$r_o$/λ) sin(θ/2) < a deve ser obedecida, onde a é um número igual ou da ordem de 1. No caso de λ ~ 1.5 Å = 0,1 nm e $r_o$ ~ 1.5 nm os ângulos de espalhamento devem ser tais que 40π sin(θ/2) < a, o que implica em termos ângulos θ muitos pequenos, no máximo da ordem de 3 ~ 4º. Isso significa que a difração deve ser efetuada a *pequenos ângulos* como nas técnicas de SANS ("Small Angle Neutron Scattering") ou SAXS ("Small Angle X−ray Scattering").[4,5]

Substituindo (3.15) na (2.3) verificamos que[15,16]

$$S(q) = 1 + (D/r_o^D) \int_0^\infty r^{D-1} \exp(-r/\xi)\, [\sin(qr)/qr]\, dr$$

$$= 1 + (1/qr_o)^D \{D\, \Gamma(D-1)/[1+1/(q\xi)^2]^{(D-1)/2}\} \sin[(D-1)\tan^{-1}(q\xi)] \quad (3.16)$$

A função S(q) é obtida integrando r a partir de r = 0, pois, o raio e o volume do monômero são considerados desprezíveis em comparação com as respectivas dimensões do agregado.

De acordo com teoria[15,16] de SAXS (ou SANS) a dimensão D é determinada analisando S(q) na região $1/\xi \ll q \ll 1/r_o$ onde $S(q) \sim q^{-D}$. Na região onde temos q muito pequenos tal que $q\xi \ll 1$ (consequentemente, $qr_o \ll 1$) temos $S(q) \approx \Gamma(D+1)(\xi/r_o)^D\{1- [D(D+1)/6]q^2\xi^2\}$ com a qual determina–se o raio de giração $R_g(D,\xi) = [D(D+1)/2]^{1/2} \xi$. Para valores grandes de q tais que $qr_o \gg 1$ a função $S(q) \to 1$. Nessas condições[15,16] a intensidade I(q) do espalhamento é dominada pela difração individual $I_o(q)$ dos monômeros dada por $I_o(q) \sim q^{(D_s - 6)}$, onde $D_s$ é a *dimensão fractal de superfície* dos monômeros.

Como exemplo vamos calcular S(q) usando a (3.16) no caso de um filme de PMMA–Au analisado[17] com SAXS para o qual obteve–se D = 1.70, $r_o$ = 1.64 nm e $\xi$ =13.08 nm. Substituindo esses valores em (3.16) calculamos S(q) em função de x = q$\sigma$ ($\sigma \sim r_o$) vista na Fig.6. Que mostra, conforme comentamos acima, uma função de estrutura S(q) que é completamente diferente das expostas nas Figuras 2–4.

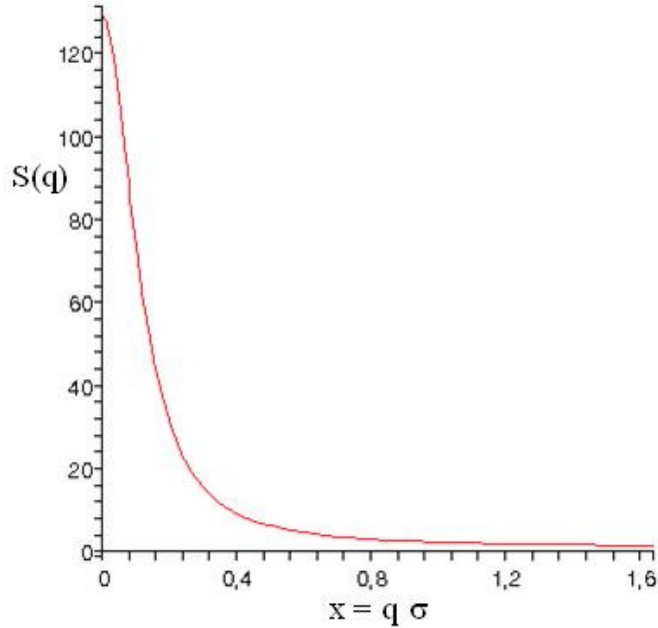

**Figura 6.** S(q) definida por (3.16) em função de x = q$\sigma$ no caso de um filme fractal de PMMA–Au analisado[16] com SAXS para o qual obteve–se D = 1.70, $r_o$ = 1.64 nm e $\xi$ =13.08 nm.

## 4.Conclusões.

Conforme Seções 1–3 se o objetivo é determinar o potencial efetivo de interação entre as partículas de um material usando, por exemplo, a difração de nêutrons ou raios–X deve–se considerar o *espalhamento a grandes ângulos*. Se a intenção é obter as propriedades fractais do material com as mesmas técnicas temos de considerar o *espalhamento a pequenos ângulos*. As funções de estrutura S(q) a grandes e a pequenos ângulos, das quais extraímos as informações necessárias, são completamente diferentes.

## APÊNDICE

A partir de (2.3) pondo q = 0 obtemos

$$S(0) = 1 + 4\pi\Phi \int_0^\infty [g(r) - 1] r^2 dr = 1 + 4\pi (N/V) \int_0^\infty [g(r) - 1] r^2 dr =$$

$$= 1 + (N/V)\{ -4\pi r_o^3/3 + \int_{r_0}^\infty [g(r) - 1] r^2 dr \}$$

$$= 1 - V_o/V + v/V = (V-V_o)/V + v/V$$

onde $v = N \int_{r_0}^\infty [g(r)-1] r^2 dr$ e $V_o = 4\pi N r_o^3/3$ = volume ocupado pelas bolinhas. A diferença $V-V_o$ é efetivamente o volume vazio que teríamos no agregado se todas as bolinhas estivessem compactadas. O volume $v$ é o "provável" volume vazio no sistema gerado pela flutuação estatística da densidade de partículas $\Phi$. Pode-se mostrar que[2] $S(0)=\Phi\chi_T kT$, onde $\chi_T = (\partial\Phi/\partial P)_{V,T}/\Phi$ é a *compressibilidade isotérmica*,

## REFERÊNCIAS